\documentstyle[pra,aps,psfig]{revtex}
\tighten
\newcommand{\bsquare}{\hbox{\rule{6pt}{6pt}}}
\newcommand{\reals}{\mbox{I$\!$R}}
\newcommand{\nums}[1]{\mbox{Z}_{#1}}
\newcommand{\ket}[1]{\left | #1 \right \rangle}
\newcommand{\bra}[1]{\left \langle #1 \right |}
\newcommand{\amp}[2]{\left \langle #1 | #2 \right \rangle}
\newcommand{\proj}[1]{\ket{#1} \bra{#1}}
\newcommand{\tr}{{\rm \, Tr }\, }
\begin{document}
\draft
\title{Accessible information and optimal strategies \\for real symmetrical
quantum sources}
\author{Masahide Sasaki$^{1}$, Stephen M. Barnett$^{2}$, Richard Jozsa$^{3}$,
Masao Osaki$^{4}$ and Osamu Hirota$^{4}$}
\address{${}^1$Communications Research Laboratory,
Ministry of Posts and Telecommunications\\
 Koganei, Tokyo 184-8795, Japan}
\address{${}^2$Department of Physics and Applied Physics, 
University of Strathclyde, \\ Glasgow G4 0NG, Scotland}
\address{${}^3$School of Mathematics and Statistics, 
University of Plymouth,\\
Plymouth, Devon PL4 8AA, England}
\address{${}^4$Research Center for Quantum Communications, 
Tamagawa University\\ Tamagawa-gakuen, Machida, Tokyo 194-8610, Japan}

\date{15 Dec 1998}

\maketitle

\begin{abstract}
We study the problem of optimizing the Shannon mutual information
for sources of real quantum states i.e. sources for which there is
a basis in which all the states have only real components. We
consider in detail the sources ${\cal E}_M$ of $M$ equiprobable
qubit states lying symmetrically around the great circle of real
states on the Bloch sphere and give a variety of explicit optimal
strategies. We also consider general real group-covariant sources for
which the group acts irreducibly on the subset of all real states
and prove the existence of a real group-covariant optimal strategy,
extending a theorem of Davies 
(E. B. Davies, IEEE. Inf. Theory {\bf IT-24}, 596 (1978)).  
Finally we propose an optical scheme
to implement our optimal strategies, simple enough to be realized 
with present technology.
\end{abstract}

\pacs{PACS numbers:03.65.Bz, 89.70.+c, 42.79.Sz, 89.80.+h, 32.80.-t}


\section{Introduction}\label{sec1}

There are two principal measures of quality in the quantum
detection problem for a given finite number of quantum states with
fixed prior probabilities. One is the minimization of a specified
Bayes cost, and the other is the maximization of the Shannon mutual
information \cite{Holevo73_condition,Helstrom_QDET,PeresBook}. The
former is useful if one has to reach a decision after performing a
single quantum measurement whereas the latter is more relevant for
the problem of transmitting as much classical information as possible
using the given ensemble of states.
In this paper we will consider the problem of maximizing
the Shannon mutual information for a certain class of quantum
ensembles.

In a general communication setting, let $\{x_i\in X\}$ be input
letters and let $\{\xi_i\}$ be their prior probabilities. Let us
denote output letters by $\{y_j\in Y\}$. Both the Bayes cost and
the Shannon mutual information are defined in terms of the
conditional probability $P(j|i)$ of obtaining output $y_j$ provided
that the letter sent was $x_i$. The former is defined as
\begin{equation}
B(X:Y)=\sum_{ij} C_{ij}\xi_i P(j|i),
\label{Bayes_cost}
\end{equation}
for a Bayes cost matrix $[C_{ij}]$, while the latter is defined as
\begin{equation}\label{ixy}
I(X:Y)=\sum_i \xi_i \sum_j P(j|i) \log \frac{P(j|i)}
{\displaystyle\sum_k \xi_k P(j|k)}.
\end{equation}
(Since all the results in this paper are valid for any logarithm
base, we shall specify the base only where necessary.)
In classical information theory, the channel matrix $[P(j\vert i)]$
is given and fixed, characterising the noise in the channel. In
contrast, in a quantum information theoretic context where signal
carriers are to be quantum states transmitted without noise, the
channel matrix generally becomes a variable. This is because the
act of quantum detection itself generally has a probabilistic
output so the channel matrix is dependent on the choice of quantum
detection strategy. More precisely, the input letters correspond to
a set of positive trace class operators of trace one
$\{\hat\rho_i\}$ on a Hilbert space ${\cal H}_s$. A quantum
detection strategy is described by a positive operator-valued
measure (POVM)  on ${\cal H}_s$. A POVM is any set $\{\hat\pi_j\}$
of hermitian positive operators forming a resolution of the
identity:
\begin{equation}
\hat\pi_j^\dagger=\hat\pi_j,\quad \hat\pi_j\geq 0\quad\forall j, \quad
\sum_j\hat\pi_j=\hat I.
\end{equation}
The detection operator $\hat\pi_j$ corresponds to the output letter
$y_j$ and the conditional probabilities are given by
\[
P(j\vert i)= {\rm Tr}(\hat\pi_j\hat\rho_i).
\]
Thus in the quantum
context the optimization of $I(X:Y)$ is carried out with respect to
the choice of POVM $\{\hat\pi_j\}$ for fixed ensemble ${\cal E} =
\{\hat\rho_i ; \xi_i\}$ (i.e. with fixed letter states $\hat\rho_i$
and fixed prior probabilities $\xi_i$). The maximum value of
$I(X:Y)$ is called the accessible information of the ensemble $\cal
E$.

The set $\cal P$ of all POVM's is a convex set and $I(X:Y)$ enjoys
the following fundamental property:
\\
{\bf (CONV)}: For a fixed
ensemble ${\cal E} =\{ \hat\rho_i ; \xi_i \} $, $I(X:Y)$ is a
convex function on $\cal P$.
\\
A proof of {\bf (CONV)} is given in
theorem 2.7.4 of \cite{covth}. Let $I({\cal E}:{\cal A})$ denote
the mutual information obtained from the POVM $\cal A$ applied to
the ensemble $\cal E$. Then if $\cal A$ is a convex combination of
POVMs ${\cal A}_i $:
\[
{\cal A} = p_1 {\cal A}_1 + \cdots p_n {\cal A}_n.
\]
it follows
from {\bf (CONV)} that
\begin{equation}
\label{conmax} I({\cal E}:
{\cal A}) \leq \sum_i p_i I({\cal E}:{\cal A}_i ) \leq \max_i
I({\cal E}:{\cal A}_i ).
\end{equation}

The Bayes cost $B(X:Y)$ is an affine concave function on the convex
set $\cal P$. Therefore the Bayes cost minimization problem is a
kind of linear programming problem and is expected to have a unique
solution. A necessary and sufficient condition for specifying the
optimum solution is known\cite{Holevo73_condition,Helstrom_QDET}.
On the other hand, the Shannon mutual information $I(X:Y)$ is a
nonlinear and convex function on $\cal P$. The maximization of this
quantity is a much harder problem and only a necessary condition
for the optimum is known \cite{Holevo73_condition}.
Thus the maximization of $I(X:Y)$ with respect to the detection strategy
$\{\hat\pi_j\}$ is a basic and open problem in quantum information
theory.

In this problem, the number of outputs is not necessarily
the same as the number of the inputs. The optimum solution is not
necessarily unique either. However it is known that there must be
at least one optimum solution which corresponds to an extreme point 
of the convex set $\cal P$. This is due to the convexity of the 
function $I(X:Y)$. Such an extreme point is a set of rank one elements, 
which means that each $\hat\pi_j$ has the form $\kappa\proj{v}$ where 
$\ket{v}$ is a pure state and $0\leq\kappa\leq 1$. 
The number of elements, $N$, can be bounded by
$d\leq N\leq d^2$ where $d$ is the dimension of the Hilbert space
${\cal H}_s$ of which the input state ensemble $\{\hat\rho_i\}$ is
made \cite{Davies78}. 
$I(X:Y)$ is also possibly maximized at some interior points of 
$\cal P$ as well. 
In that case the number of outcomes may exceed $d^2$. 
Explicit examples of optimal solutions have been given for
binary ensembles \cite{FuchsPeres96,Ban97_C1,Osaki97_C1} and for
the ensemble of four qubit states with tetrahedral symmetry
\cite{Davies78}. The latter is a specific example of a general
result of Davies \cite{Davies78} characterising the form of an
optimal strategy for any symmetrical ensemble whose symmetry group
acts {\it irreducibly} on the {\it whole} state space.

In this paper we will study the accessible information and
corresponding optimal strategies for an ensemble ${\cal E}_M$ of
$M$ qubit states with symmetry group $\nums{M}$, the group of
integers modulo $M$. Some of our results will also apply to more
general ensembles. ${\cal E}_M$ may be explicitly described as
follows.
Let
$\{ \left(
\begin{array}{c} 1 \\ 0 \end{array} \right) ,\left(
\begin{array}{c} 0 \\ 1
\end{array} \right) \} $ be the $z$-spin eigenstates and write $\ket{\psi_0}
= \left( \begin{array}{c}1 \\ 0 \end{array}
\right)$.
Let
\begin{equation}\label{zm}
\hat V\equiv{\rm exp}(-i{\pi\over M}\hat\sigma_y)
=\left(
\begin{array}{cc}
\cos {{\pi}\over M} & -\sin {{\pi}\over M}  \\
\sin {{\pi}\over M} & \cos {{\pi}\over M}
\end{array}
\right).
\end{equation}
Then ${\cal E}_M$ consists of the $M$ states
\begin{equation}
\label{psik} \ket{\psi_k} = \hat V^k \ket{\psi_0}
= \left( \begin{array}{c} \cos \frac{k\pi}{M} \\ \sin \frac{k\pi}{M}
\end{array} \right),
\quad k=0, \ldots , M-1,
\end{equation}
taken with equal prior probabilities $\xi_k = \frac{1}{M}$. Note that
these states (in the $z$-spin basis) involve only {\it real}
components. On the Bloch sphere they are equally spaced around a
great circle $C$ in the $x-z$ plane consisting of all real states.
The antipodal points which have $C$ as equator, are the two
$\hat\sigma_y$ eigenstates. Thus ${\cal E}_M$ is clearly
symmetrical with respect to the group $\nums{M}$ whose generator is
represented by ${2\pi \over M}$ rotation about the axis joining the
$\hat\sigma_y$ eigenstates. At the Hilbert space level the
operators $\hat{V}^k$ in Eq. (\ref{zm}) provide a projective
unitary representation of $\nums{M}$ (e.g. $\hat{V}^M =-I$ and c.f.
Eq. (\ref{rep}) later).

This symmetry group does not act irreducibly on the whole state
space. Indeed the $\hat\sigma_y$ eigenstates are left invariant by
the group action. (Irreducibility on the whole state space requires
that the only invariant point is the maximally mixed state ${1\over
2}\hat I$.) Hence we cannot apply Davies' theorem \cite{Davies78} to
provide an optimal strategy for ${\cal E}_M$. Nevertheless we will
prove that the conclusion of Davies' theorem remains true in this
case i.e. that there exists a pure state $\ket{a_0}$ such that the
$\nums{M}$-symmetric POVM
\[
{\cal A}_M = \{ {2\over M}
\proj{a_k} : k=0, \ldots ,M-1 \} \quad \mbox{where $\ket{a_k}=
\hat{V}^k \ket{a_0}$},
\]
is an optimal strategy for ${\cal E}{_M}$. Furthermore we will show
that $\ket{a_0}$ may be taken to be the state orthogonal to
$\ket{\psi_0}$.

The case $M=3$ is of particular interest. It is the so-called trine
ensemble which has been much studied
\cite{Holevo73_POM,Hausladen_SubOptMeas95,pw92}. Holevo in 1973
\cite{Holevo73_POM} showed that no von Neumann measurement in
${\cal H}_2$ can be an optimal strategy, demonstrating the
necessity of considering more general POVMs in quantum detection
theory. Since that time it has been conjectured that the strategy
${\cal A}_3$ above is optimal for the trine source. Our results
resolve this conjecture affirmatively.

The strategy ${\cal A}_M$ has $M$ elements. However, as noted
above, for ensembles in $d=2$ dimensions there is always an optimal
strategy with at most $d^2 =4$ elements (which does not increase
with $M$). We will show that the ensembles ${\cal E}_M$ always have
an optimal strategy with at most 3 elements and explicit strategies
of this form will be described for all $M$. If $M$ is even then
${\cal E}_M$ consists of ${M \over 2}$ pairs of orthogonal states.
Let $\{
\ket{\xi}, \ket{\eta} \}$ be any one of these pairs. We will show
that the two-element POVM $\{ \proj{\xi}, \proj{\eta} \}$ (a
regular von Neumann measurement) is always an optimal strategy when
$M$ is even. We will also describe further optimal $K$-element
POVMs where $K$ lies between 3 and $M$.

\section{A Group-theoretic Approach}\label{sec2new}

We begin by setting up a group-theoretic formalism for symmetric
ensembles, leading to a main result (theorem 1) which applies to
symmetric ensembles of real states in any dimension $d\geq 2$. An
essential requirement in many of our results will be that various
states and unitary operators be {\it real}. The requirement that a
state or operator be real has of course, no intrinsic physical
meaning. When we speak of real states and real operators we will
always mean simply that {\it there exists a basis of the Hilbert
space relative to which all the required objects simultaneously
have real components or real matrix elements}.

A projective unitary representation of a group $G$ is an assignment
of a unitary operation $\hat U_g$ to each member of $G$ satisfying
\begin{equation}
\label{rep} \hat U_{g_1}\hat U_{g_2}
= e^{i\phi (g_1 , g_2 )} \hat U_{g_1 g_2},
\end{equation}
where the phases $\phi (g_1 , g_2 )$ may be chosen arbitrarily.
A finite ensemble ${\cal E}$ of equiprobable (generally mixed)
states is said to be symmetric with respect to the group $G$, or
$G$-covariant, if the following condition is satisfied: there is a
projective unitary representation
$\{ \hat U_g \}$ of $G$ such that for
all $g$, $\hat U_g \hat \rho \hat U_g^\dagger$ is in $\cal E$
whenever $\hat \rho$ is
in $\cal E$. We write
\begin{equation}
\label{action} g\hat \rho  = \hat U_g
\hat \rho \hat U_g^\dagger,
\end{equation}
for the action of $g$ on the state
$\hat \rho$. The phases $\phi (g_1 , g_2 )$ do not appear in Eq.
(\ref{action}) and $g_1(g_2 (\hat\rho))=(g_1 g_2)(\hat\rho)$.
Note that, in contrast to Davies \cite{Davies78} we do not require
that $G$ parameterises $\cal E$ i.e. $G$ need not act transitively
on the set of states of $\cal E$. For example, ${\cal E}_M$ is
$\nums{M}$-covariant and the action is transitive, but ${\cal
E}_{2N}$ is also $\nums{2}$- and $\nums{N}$-covariant via
non-transitive actions.

A $G$-covariant POVM $\cal A$ (for the projective unitary
representation $\{\hat U_g \}$) is a POVM such that
$\hat U_g \hat A \hat U_g^\dagger $
is in $\cal A$ whenever $\hat A$ is in $\cal A$.
We write
\begin{equation}
\label{dual}
g\hat A\equiv\hat U_g \hat A \hat U_g^\dagger,
\end{equation}
for the action of $g$ on a POVM element $\hat A$. From Eqs.
(\ref{action}) and (\ref{dual}) we see that
$ \tr (\hat A\hat \rho )= \tr (g\hat A\cdot g\hat \rho )$
i.e. the probability of outcome $\hat A$ on state $\hat \rho$ is
$G$-invariant. Hence
\begin{equation}
\label{trg}
\tr (g\hat A\cdot\hat\rho)=\tr(\hat A\cdot g^{-1}\hat\rho),
\end{equation}
so that the set of probabilities of the $G$-shifted outputs
$g\hat A$ on a fixed input $\hat \rho$ are obtained as
a permutation of the set of probabilities of the unshifted
output $\hat A$ acting on suitably shifted inputs.

Let $\cal E$ be a $G$-covariant ensemble with projective unitary
representation $\{ \hat U_g \}$. We aim to find conditions on
$\{ \hat U_g \}$ which will guarantee the existence of a
$G$-covariant POVM
${\cal A} = \{ \hat A_g : g\in G \}$ with elements parameterised
by $G$, and having group action $g\hat A_h =\hat A_{gh}$.
Thus if $e$ is the identity of $G$ we have
\begin{equation}
\label{ag} \hat A_g = \hat U_g \hat A_e \hat U_g^\dagger,
\end{equation}
and we require
\begin{equation}
\label{ident} \hat M\equiv  \sum_{g\in G}
\hat A_g = \hat I.
\end{equation}
(Later we will take the elements of $\cal A$
to be rank 1 and consider the question of when $\cal A$ is an
optimal strategy for $\cal E$.) From Eq. (\ref{ag}) we see that
$\hat M$ commutes with all the $\hat U_g$'s:
\begin{equation}
\label{inv} \hat U_g \hat M = \hat M\hat U_g.
\end{equation}
Thus if the set $\{ \hat U_g \}$ acts irreducibly on the state space
(i.e. there is no proper invariant subspace) the Schur's lemma will
guarantee that Eq. (\ref{ident}) holds. This fact is used by Davies
\cite{Davies78} to characterise an optimal strategy for any
$G$-covariant ensemble whose symmetry group acts irreducibly on the
whole state space. However this condition of full irreducibility on the
whole state space is not necessary for Eq. (\ref{ident}) to hold.
We will use the following more general form of Schur's lemma:

{\bf Lemma 1}: Let $\{ \hat M_g \}$ be any set of non-singular
$d$ by $d$ matrices over some field $F$ which acts irreducibly
on the vector space $V=F^d$
(i.e. there is no proper subspace mapped to
itself by all the $\hat M_g$'s).
Suppose that $\hat K$ is any matrix that
commutes with all the $\hat M_g$'s:
\begin{equation}
\label{comm} \hat K\hat M_g = \hat M_g \hat K.
\end{equation}
Then:\\(a) either $\hat K=0$ or $\hat K$ is non-singular,\\(b)
If $\hat K$ has a
non-zero eigenvalue $\lambda$ in $F$,
then $\hat K=\lambda \hat I$.

{\bf Proof}: (a) Let $\hat K(V)$ denote the image of $V$ under
the map $\hat K$ and similarly for $\hat M_g (V)$.
Since $\hat M_g$ is non-singular we
have $\hat M_g (V)=V$. By Eq. (\ref{comm}) we have
$\hat M_g \hat K(V)= \hat K\hat M_g (V)
=\hat K(V)$ i.e. $\hat K(V)$ is an invariant subspace.
Hence either $\hat K(V)=0$ (in which case $\hat K=0$) or else
$\hat K(V)=V$ (in which case $\hat K$
is non-singular).\\(b) If $\hat K$ has eigenvalue $\lambda$ in $F$
then $\hat B=\hat K-\lambda \hat I$ is singular.
Also $\hat B\hat M_g = \hat M_g \hat B$ for all $g$. Hence
by (a), $\hat B$ must be zero i.e. $\hat K=\lambda \hat I$.
$\bsquare$

We will apply this lemma with $F=\reals$ to obtain useful results
about $G$-covariant ensembles of {\it real} states whose group $G$
acts irreducibly only on the restricted set $\reals^d$ of real
states (but not necessarily irreducibly on the full state space).
This is the case for our ensembles ${\cal E}_M$. Let $|G|$ denote
the size of $G$ and let $d=\tr \hat I$ be the dimension of
the Hilbert space.

{\bf Lemma 2}:
Suppose that $\{ \hat U_g \}$ is a projective unitary
representation of $G$ such that $\hat U_g$ are all
{\it real} matrices
and $\{ \hat U_g \}$ acts irreducibly on $\reals^d$.
Let $\ket{v}\in
\reals^d$ be any real state.
Write
\[
\hat A_g = {d\over |G|} \hat U_g \proj{v} \hat U_g^\dagger.
\]
Then
$\{
\hat A_g : g\in G \}$ is a $G$-covariant POVM i.e.
$\sum_{g\in G} \hat A_g=\hat I$.

{\bf Proof}: Let $\hat M= \sum_{g\in G} \hat A_g$.
Then $\hat M$ is a real matrix
and $\hat M\hat U_g = \hat U_g \hat M$ for all $g\in G$.
Also $\hat M$ is a hermitian
positive matrix (being a sum of projectors with positive
coefficients) so it has a real positive eigenvalue $\lambda>0$.
By the previous lemma, $\hat M=\lambda\hat I$. Since
$\tr \hat A_g = {d\over |G|}$ for all $g$,
we get $\tr \hat M = d = \tr \hat I$ so $\lambda=1$.
$\bsquare$

{\bf Theorem 1}: Let $\cal E$ be any ensemble of equiprobable real
states in dimension $d$. Suppose that $\cal E$ is $G$-covariant
with respect to a projective unitary representation
$\{ \hat U_g \}$ of
real matrices which acts irreducibly on $\reals^d$. Then there
exists a real pure state $\ket{v}$ such that the $G$-covariant POVM
${\cal D} = \{ \hat D_g : g\in G\}$ defined by
\[ \hat D_g = {d\over |G|} \hat U_g \proj{v} \hat U_g^\dagger, \]
is an optimal strategy for $\cal E$.

{\bf Proof}: We will work in the basis with respect to which the
states of $\cal E$ and the matrices $\hat U_g$ have real entries.
Let ${\cal A}= \{ \hat A_1 , \ldots ,\hat A_n \}$ be {\it any}
optimal POVM for
$\cal E$. We will transmogrify $\cal A$ into the required form
while preserving optimality. First strip off all imaginary parts of
the entries of the matrices $\hat A_k$.
Let $\tilde{A}_k = Re (\hat A_k )$
and $\tilde{\cal A} = \{ \tilde{A}_1 , \ldots , \tilde{A}_n \}$.
Then $\tilde{\cal A}$ is again a POVM and has real symmetric
matrices as elements. (To see that $\tilde{A}_k$ is a positive
matrix note that $A_k$ positive implies that the complex conjugate
$\hat A^*_k$ is positive so $\tilde{A}_k
= {1\over 2} ( \hat A_k + \hat A^*_k )$
must be positive. Also $\sum \hat A_k = \hat I$ and $\hat I$
is real so $\sum
\tilde{A}_k = I$ too.) Next note that
$\tr \hat A_k \hat \rho = \tr\tilde{A}_k \hat \rho$
for any real state $\hat \rho$ (since $Im(\hat A_k )$ is
antisymmetric) so $\tilde{\cal A}$ remains an optimal strategy.

In general $\tilde{\cal A}$ will not have rank 1 elements even if
$\cal A$ had rank 1 elements. Thus decompose each $\tilde{A}_k$
into its rank 1 eigenprojectors (multiplied by the corresponding
eigenvalues) which are necessarily real as the eigenvalues/vectors
of any real symmetric matrix are real. Then form the larger POVM
${\cal B} = \{ \hat B_1 , \ldots , \hat B_m \}$
comprising all the scaled
rank 1 eigenprojectors above. Such a refinement of a POVM can never
decrease the mutual information so $\cal B$ with real rank 1
elements, is still optimal.

Now look at
\begin{equation}
\label{cig} \hat C_{kg} = {1\over |G|} g\hat B_k
\quad \mbox{for $g\in G$ and $k=1, \ldots , m$}.
\end{equation}
Note that $\sum_{kg} \hat C_{kg}=\hat I$ since
$\sum \hat B_k =\hat I$ and $g\hat I=\hat I$ for
all $g$. Let ${\cal C} =\{ \hat C_{kg} \}$ be the corresponding
POVM with $|G|m$ elements. Thus $\cal C$ is $G$-covariant but
the action of $G$ is not transitive.
We finally aim to cut down $\cal C$ to a
smaller optimal $G$-covariant POVM with elements labeled by $G$.

Let $I({\cal E}:{\cal A})$ denote the mutual information obtained
from any POVM $\cal A$ applied to any ensemble $\cal E$. First we
show that $I({\cal E}:{\cal C})= I({\cal E}:{\cal B})$ so that
$\cal C$ remains optimal. Let us label the inputs by $i\in {\cal
I}$ and denote conditional probabilities for $\cal C$ by $P(kg|i)$.
Denote the conditional probabilities for $\cal B$ by $P_B (k|i)$
and let $\xi$ be the constant prior input probability.
Then
\[
P(kg|i)
=\tr \hat C_{kg}\hat \rho_i
={1\over |G|} \tr g\hat B_k\cdot\hat \rho_i.
\]
According to
Eq. (\ref{trg}), for each fixed $g$ and $k$ the resulting set of
probabilities labelled by $i\in {\cal I}$, will be just a {\it
permutation} of the set $P_B (k|i)$, rescaled by ${1\over |G|}$.
Thus
\[ \sum_l \xi P(kg|l) = {1\over |G|} \sum_l \xi P_B (k|l), \]
will be independent of $g$ and also
\[ \sum_i P(kg|i) \log {P(kg|i) \over \xi \sum_l P(kg|l)} =
{1\over |G|} \sum_i P_B (k|i) \log { P_B (k|i) \over
\sum_l \xi P_B (k|l)}, \]
will be independent of $g$. The mutual informations $I({\cal
E}:{\cal C})$ and $I({\cal E}:{\cal B})$ are given by (c.f. Eq.
(\ref{ixy})):
\[ I({\cal E}:{\cal C}) = \sum_i \xi \sum_{kg} P(kg|i) \log
{P(kg|i) \over \xi \sum_l P(kg|l) },  \]
\[
I({\cal E}:{\cal B}) = \sum_i \xi \sum_{k} P_B (k|i) \log
{P_B (k|i) \over \xi \sum_l P_B (k|l) }.
\]
On substituting the
above $G$-invariant expressions into $I({\cal E}:{\cal C})$ we
readily get $I({\cal E}:{\cal C})= I({\cal E}:{\cal B})$. (Our
argument is actually an explicit example of the claim in lemma 5 of
\cite{Davies78}). Hence $\cal C$ remains optimal.

Finally note that for each $i$, $\hat B_i / (\tr \hat B_i )$
is a real pure state so by lemma 2,
\[
{\cal D}_i = \left\{ {d\over |G|} {g\hat B_i
\over \tr \hat B_i } : g\in G \right\},
\]
is a POVM for each $i$. Now
${ \tr \hat B_i \over d} {\cal D}_i = \{
{1\over |G| }g\hat B_i : g\in G \}$
so $\cal C$ is a convex combination
\[ {\cal C} = \sum_{i=1}^m {\tr \hat B_i \over d } {\cal D}_i. \]
Hence by Eq. (\ref{conmax})
\[ I({\cal E}:{\cal C}) \leq \max_i I({\cal E}:{\cal D}_i ). \]
Since $\cal C$ was optimal it follows that at least one of the
${\cal D}_i$'s is optimal. This gives an optimal $G$-covariant POVM
with real rank 1 elements, parameterised by $G$, completing the
proof.
$\bsquare$

\section{Optimal Strategies for ${\cal E}_M$} \label{optem}

We now return to the $\nums{M}$-covariant ensemble ${\cal E}_M$ in
2 dimensions, comprising the states
\[ \ket{\psi_k} = \left(
\begin{array}{c}
\cos {k\pi \over M} \\ \sin {k\pi \over M}
\end{array}
\right), \quad k=0, \ldots , M-1,
\]
with equal prior probabilities ${1\over M }$. According to theorem
1, there must exist an optimal $\nums{M}$-covariant POVM
${\cal A}= \{ \hat A_0 , \ldots ,\hat A_{M-1} \}$
with $M$ real rank 1 elements.
The elements will have the form $\hat A_j = \proj{a_j}$ with
\begin{equation}
\label{povmth}
\ket{a_j} =\hat{V}^j \ket{a_0} =
\sqrt{2\over M}
\left( \begin{array}{c} \cos (\theta +{j\pi \over M}) \\
\sin (\theta + {j\pi \over M}) \end{array} \right),
\quad j=0, \ldots , M-1,
\end{equation}
and $\hat{V}$ is given in Eq. (\ref{zm}). The conditional
probabilities $p(j|k)= |\amp{a_j}{\psi_k}|^2$ may be readily
computed and after some rearrangement we obtain the mutual
information $I(\theta )$ explicitly as
\begin{equation}
\label{mutinf}
I(\theta ) =
{1\over M} \sum_{k=0}^{M-1} (1+ \cos (2\theta-{2k\pi \over M}))
\log (1+ \cos (2\theta-{2k\pi \over M})).
\end{equation}
In this section, the base of the logarithm is taken as $e$. (For
this base the numerical value of Eq. (\ref{mutinf}) is the amount
of information in nats rather than bits.) From the symmetry,
$I(\theta )$ is a periodic function with period $\pi
\over M$. Figure 1 shows numerical plots of $I(\theta )$ for
$M=2,3,4$ and 5 and illustrates the following basic property:

{\bf Lemma 3}: For each $M$, $I(\theta )$ has a global maximum at
$\theta
= {\pi
\over 2}$.

{\bf Proof}: Since $\vert \cos (2\theta-{{2k\pi}\over M})\vert<1$,
$I(\theta) $ can be expanded by using the formula
\begin{equation}
(1+x)\ln(1+x)=x+\sum_{n=2}^\infty
{{(-1)^n}\over{n(n-1)}} x^n, \quad \vert x\vert<1,
\end{equation}
We get
\begin{eqnarray}
I(\theta)
&=&
{1\over M}\sum_{k=0}^{M-1}
\Big[
{\rm cos}(2\theta-{{2k\pi}\over M}) +  \sum_{n=2}^\infty
    {{(-1)^n}\over{n(n-1)}}  {\rm cos}^n
(2\theta-{{2k\pi}\over M})
\Big]
\nonumber \\
&=& {1\over M} \sum_{n=2}^\infty  {{(-1)^n}\over{n(n-1)}}
           \sum_{k=0}^{M-1} {\rm cos}^n (2\theta-{{2k\pi}\over M}).
\nonumber
\end{eqnarray}
since $\sum_{k=0}^{M-1} {\rm cos}(2\theta-{{2k\pi}\over M}) =0$.
Next we separate out the even and odd parts of the series and
replace powers of cosines by multiple angle cosines to get:
\begin{eqnarray} I(\theta)
&=& {1\over M} \sum_{n=1}^\infty  {{(-1)^{2n}}\over{2n(2n-1)}}
                 \sum_{k=0}^{M-1} {\rm cos}^{2n} (2\theta-{{2k\pi}\over M})
\nonumber \\
&+& {1\over M} \sum_{n=1}^\infty  {{(-1)^{2n+1}}\over{(2n+1)2n}}
                 \sum_{k=0}^{M-1} {\rm cos}^{2n+1} (2\theta-{{2k\pi}\over
M})
\nonumber \\
&=& {1\over M} \sum_{n=1}^\infty  {{(-1)^{2n}}\over{2n(2n-1)}}
                 \sum_{k=0}^{M-1} {1\over{2^{2n-1}}}
                 \Big\{
                          {1\over2} \left(\begin{array}{c} 2n \cr n
\end{array}\right)
                       + \sum_{l=0}^{n-1}
                          \left(\begin{array}{c} 2n \cr l
\end{array}\right)
                         {\rm cos} \left[(2n-2l)(2\theta-{{2k\pi}\over
M})\right]
                 \Big\}
\nonumber \\
&+& {1\over M} \sum_{n=1}^\infty  {{(-1)^{2n+1}}\over{(2n+1)2n}}
                 \sum_{k=0}^{M-1} {1\over{2^{2n}}}
                 \Big\{ \sum_{l=0}^{n}
                          \left(\begin{array}{c} 2n+1 \cr l
\end{array}\right)
                         {\rm cos} \left[(2n+1-2l)(2\theta
                         -{{2k\pi}\over M})\right]
         \Big\}
          ,\label{eqn:I_max_proof1}
\end{eqnarray}

\noindent
Then recall that
\begin{equation}
\sum_{k=0}^{M-1} {\rm cos} \left(L{{2k\pi}\over M}\right)
= \left \{
\begin{array}{rl}
M  & \quad \mbox{for } {L/M}=q \mbox{ (integer)} \\
0  & \quad \mbox{for } {L/M}\ne\mbox{ integer}
\label{eqn:k_sum_delta_function}
\end{array}
\right. .
\end{equation}

\noindent
Applying this to Eq. (\ref{eqn:I_max_proof1}) with $L=2n-2l$ and
$L=2n+1-2l$ in the even and odd series, we get:
\begin{eqnarray}
I(\theta)
&=&
\sum_{n=1}^\infty {{(-1)^{2n}}\over{2n(2n-1)2^{2n-1}}}
 \Big[ {1\over2}\left(\begin{array}{c} 2n \cr n \end{array}\right)
      + \sum_{l=0}^{n-1}\sum_{q=0}^\infty
                \left(\begin{array}{c} 2n \cr l \end{array}\right)
                {\rm cos}(2\theta qM) \delta_{2n-2l,qM}
 \Big]
\nonumber \\
&+&
\sum_{n=1}^\infty { {(-1)^{2n+1}}\over{ (2n+1) 2n 2^{2n} } }
 \Big[ \sum_{l=0}^{n}\sum_{q=0}^\infty
       \left(\begin{array}{c} 2n+1 \cr l \end{array}\right)
       {\rm cos}(2\theta qM) \delta_{2n+1-2l,qM}
 \Big]
\nonumber \\
&=&
\sum_{n=1}^\infty
         {1\over{2n(2n-1)2^{2n}}}
         \left(\begin{array}{c} 2n \cr n \end{array}\right)
+ \sum_{q=0}^\infty f(qM) (-1)^{qM} {\rm cos}(2\theta qM)
,
\label{eqn:I_max_proof2}
\end{eqnarray}

\noindent
where
\begin{equation}
f(qM) = \sum_{n=1}^\infty \sum_{l=0}^{n-1}
{
    { \left(\begin{array}{c} 2l+qM \cr l
\end{array}\right) }
    \over
    {(2l+qM)(2l+qM-1)2^{2l+qM-1}}
}
\left(  \delta_{2n-2l,qM}+\delta_{2n+1-2l,qM}  \right).
\end{equation}
Since $f(qM)>0$,  $I(\theta)$ is maximized when $(-1)^{qM} {\rm
cos}(2\theta qM)=1$, that is, $\theta={\pi\over2}$ for all $M$.
$\bsquare$

\begin{figure}[htb]
\centerline{\psfig{file=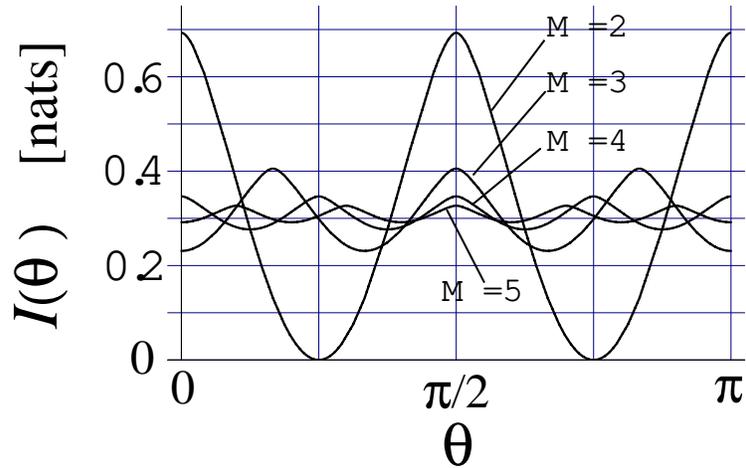,width=10cm}}
\caption{The Shannon mutual information $I(\theta)$ in nats versus
the optimization parameter $\theta$ for $M=$2, 3, 4 and 5.}
\label{I_pure}
\end{figure}

Hence in general an optimal strategy for ${\cal E}_M$ consists of
choosing a real rank 1 POVM with elements $\hat A_k$ lying in directions
orthogonal to the input states $\ket{\psi_k}$. This POVM will be
denoted by ${\cal A}_M$. The output $\hat A_k$ signifies with certainty
that the input was not $\ket{\psi_k}$ but leaves a residual
uncertainty in the remaining signal states.

For a given ensemble $\cal E$ the optimal strategy is not unique
and in practice it may be of interest to find optimal POVMs with
the minimum number of elements. The $G$-covariant optimal POVM
above has $M$ elements and we note here some ways of reducing this
number using the group theoretic approach. In the next section, by
different methods, we will show that 3 elements always suffice for
any real qubit source, and develop corresponding strategies for the
${\cal E}_M$'s.

{\bf Lemma 4}: Suppose that $k\neq 1$ divides $M$ exactly. Then
there is a $\nums{k}$-covariant optimal POVM for ${\cal E}_M$ with
$k$ real rank 1 elements.

{\bf Proof}: Since $k$ divides $M$, $\nums{M}$ has a subgroup
isomorphic to $\nums{k}$ and so ${\cal E}_M$ is
$\nums{k}$-covariant. Since $k\neq 1$, the action of $\nums{k}$
contains a non-trivial rotation so it acts irreducibly on
$\reals^2$. Thus theorem 1 immediately gives the required
result.$\bsquare$

{\bf Remark}: Lemma 4 may also be obtained by a convexity argument
as follows. We will illustrate the idea with the specific example
of $M=15$ and $k=3$. The general case is a straightforward
generalisation. $\nums{15} = \{ 0,1, \ldots ,14 \}$ has the
subgroup $\{ 0,5,10 \}$ isomorphic to $\nums{3}$.
Let ${\cal A}_{15}
= \{ \hat A_0 ,\hat A_1 , \ldots , \hat A_{14} \}$
be the optimal strategy given by theorem 1 and lemma 3,
with the direction of $\hat A_k$ being
orthogonal to the $k^{\rm th}$ state of ${\cal E}_{15}$. According
to lemma 2, the three directions 0,5,10 corresponding to the
subgroup, may be used to define a POVM. We just need to rescale
$\hat A_0 ,\hat A_5$ and $\hat A_{10}$
so that they add up to $\hat I$. The scaling
factor is ${M\over k} =5$.
Thus ${\cal B}_0 = \{ 5\hat A_0 ,5\hat A_5 ,
5\hat A_{10} \}$ is a POVM.
Now $\hat I$ is always $G$ invariant so we can
apply the group elements $l=1,2,3$ and 4 of $\nums{15}$ to ${\cal
B}_0$ to obtain POVMs
\[
{\cal B}_l \equiv l{\cal B}_0 =
\{ 5\hat A_l ,5\hat A_{l+5} ,5\hat A_{l+10} \}  
\quad \mbox{for $l=0,1,2,3,4$}.
\] Note that the ${\cal B}_l$'s have elements parameterised by the
{\it cosets} of $\nums{3}$ in $\nums{15}$.  Also by symmetry of the
construction, $I({\cal E}_{15} : {\cal B}_l )$ is independent of
$l$. Furthermore ${\cal A}_{15}$ is a uniform convex combination of
the ${\cal B}_l$'s
\[
{\cal A}_{15} = \sum_{l=0}^{4} {1\over 5}
{\cal B}_l,
\]
so by Eq. (\ref{conmax}):
\[
I({\cal E}_{15} : {\cal A}_{15} ) \leq
\max_l I({\cal E}_{15} : {\cal B}_l ).
\]
Since ${\cal A}_{15}$ was optimal we see that ${\cal B}_l$ is
optimal for each $l$. This gives the result of lemma 4 and also
identifies the directions of the $k$ element POVM as being any
chosen symmetrical set of $k$ directions orthogonal to
corresponding states of ${\cal E}_M$.$\bsquare$

An immediate special case is:

{\bf Corollary}: If $M$ is even then ${\cal E}_M$ is made up of
${M\over 2}$ pairs of orthogonal states. The von Neumann
measurement defined by any one of these orthogonal pairs is an
optimal strategy for ${\cal E}_M$.$\bsquare$

Thus if $M$ is composite we can significantly reduce the number of
elements in our optimal strategy but if $M$ is prime then this
number remains large. In the next section we give a different
approach to reducing the number of elements, showing that just 3
elements always suffices for any ensemble of real qubit states.

\section{Optimal POVMs with 3 Elements}\label{povm_w}

Davies \cite{Davies78} has shown that any ensemble in $d$
dimensions has an optimal strategy with $N$ elements where $d\leq
N\leq d^2$. This is directly based on {\bf (CONV)}, that is,
$I(X:Y)$ is a convex function on the convex set ${\cal P}$ of all
POVMs. Because of this, $I(X:Y)$ will always take its maximum value
at an extreme point of the convex set ${\cal P}$ (and also possibly
at some interior points as well). Each extreme point of ${\cal P}$
consists of $N$ rank 1 elements bounded by $d\leq N\leq d^2$. If we
restrict attention to only {\it real} ensembles then this upper
bound on $N$ can be improved as follows
\cite{Holevo98_private_commun}.

{\bf Lemma 5}: Let $\cal E$ be any ensemble of real states in $d$
dimensions. Then the Shannon mutual information can be maximized by
a POVM with $N$ elements where $d\le N\le d(d+1)/2$.

{\bf Proof}: The proof proceeds along the same lines as the
original one in ref. \cite{Davies78} with a slight replacement. For
any POVM $\{ \hat\pi_j \}$ write $\hat\pi_j=\mu_j d \bar\pi_j$
where ${\rm Tr}\bar\pi_j=1$ so
\begin{equation}
\sum_j \mu_j \bar\pi_j=\hat I/d, \quad \sum_j \mu_j=1.
\end{equation}
Let $\cal X$ be the (compact convex) set of all positive hermitian
operators with trace 1 (such as the $\bar\pi_j$'s). Since $I({\cal
E}:{\cal A})$ is a convex function on the set $\cal P$ of all POVMs
its maximum is attained at an extreme point of $\cal P$. The
essential point of the original proof in ref. \cite{Davies78} is
that every extreme point of $\cal P$  has $D+1$ rank 1 elements
where $D$ is the real dimension of $\cal X$. In the case of general
ensembles $D=d^2
-1$. In our case of real ensembles the members of $\cal X$ and $\cal P$
can be restricted to real matrices so  $\cal X$ comprises real
symmetric trace 1 matrices and $D={ d(d+1)\over 2} -1$. Hence the
extreme points of $\cal P$ have $\leq { d(d+1) \over 2}$
elements.$\bsquare$

Thus for the real ensembles ${\cal E}_M$ with $d=2$, POVMs with
three real elements suffice to provide an optimal strategy. To
describe such a POVM, we first introduce the three real
(un-normalised) vectors
\begin{mathletters}
\begin{eqnarray}
\vert\omega_0\rangle
  &=& c \left(  \begin{array}{c} 1 \cr
                                          0
              \end{array}  \right), \\
\vert\omega_1\rangle
   &=&a \left(  \begin{array}{c} {\rm cos}\varphi_a \cr
                                            {\rm sin}\varphi_a
                 \end{array}  \right), \\
\vert\omega_2\rangle
   &=&b \left(  \begin{array}{c} {\rm cos}\varphi_b \cr
                                            {\rm sin}\varphi_b
                 \end{array}  \right),
\end{eqnarray}
\end{mathletters}

\noindent
where the first vector lies along the first basis direction and the
remaining two are in general position. Imposing the  condition
$\sum_j\vert\omega_j\rangle\langle\omega_j\vert=\hat I$ we get
\begin{mathletters}
\begin{eqnarray}
c & = & \sqrt{2-a^2 -b^2 },  \\
 a^2&=&{{{\rm cos}\varphi_b}\over
            {{\rm sin}\varphi_a{\rm sin}(\varphi_a-\varphi_b)}}, \\
b^2&=&{{{\rm cos}\varphi_a}\over
            {{\rm sin}\varphi_b{\rm sin}(\varphi_b-\varphi_a)}},
\end{eqnarray}
and
\begin{equation}
0\le a^2+b^2\le2.
\end{equation}
\label{condition_phi}
\end{mathletters}

\noindent
Once the angles $\varphi_a$ and $\varphi_b$ have been chosen, $a$,
$b$ and $c$ are fixed. Finally we rotate these vectors  around the
$y$-axis through an angle $\theta$ to make the general POVM with
three real rank 1 elements:
\begin{mathletters}
\begin{equation}
\hat\omega_j(\theta)\equiv
\vert\omega_j(\theta)\rangle\langle\omega_j(\theta)\vert,
\end{equation}
\begin{equation}
\vert\omega_j(\theta)\rangle
  =\hat V(\theta)\vert\omega_j\rangle, \quad
\hat V(\theta)\equiv {\rm exp}(-i\theta\hat\sigma_y).
\end{equation}
\label{POVM_omega}
\end{mathletters}

\noindent
This gives the most general POVM $\{\hat\omega_0(\theta),
\hat\omega_1(\theta),
\hat\omega_2(\theta)\}$ in terms of three independent parameters
$\varphi_a$, $\varphi_b$
and $\theta$.

We are now in a position to maximize the Shannon mutual information
of ${\cal E}_M$ with (at most) three-element POVMs. We first give a
useful preliminary lemma.

{\bf Lemma 6:} Let ${\cal A} = \{ \lambda_a^2 \proj{a} \}$ be any
POVM with rank 1 elements labelled by $a$ where $0 < \lambda_a \leq
1$ is real and
\[ \ket{a}= \left( \begin{array}{c} \cos \theta_a \\ \sin \theta_a
\end{array} \right) \]
in the $z$-spin basis. Then the mutual information for ${\cal E}_M$
is given by
\begin{equation} \label{suma}
I({\cal E}_M : {\cal A} )= \sum_a {\lambda_a^2 \over 2} I(\theta_a
)
\end{equation}
where $I(\theta )$ is the function given in Eq. (\ref{mutinf}).

{\bf Proof:} The states $\ket{\psi_k}$ of ${\cal E}_M$ given in eq.
(\ref{psik}) lead to the conditional probabilities
\[ P(a|k)= \lambda_a^2 |\amp{\psi_k}{a}|^2 =
{ \lambda_a^2 \over 2 } ( 1+\cos (2\theta_a -{ 2k\pi \over M })) \]
Substituting these into eq. (\ref{ixy}) readily yields the formula
eq. (\ref{suma}) after a little algebra.$\bsquare$

{\bf Theorem 2}: The Shannon mutual information of ${\cal E}_M$
(for $M>2$) is maximized by the POVM ${\cal W}=\{
\hat\omega_j^\ast=\vert\omega_j^\ast\rangle\langle\omega_j^\ast\vert :
j=0,1,2 \} $ where
\begin{mathletters}
\begin{eqnarray}
\vert\omega_0^\ast\rangle
  &=&\left(  \begin{array}{c} 0 \cr
                                             \sqrt{2-a^2-b^2}
                   \end{array}  \right), \quad \\
\vert\omega_1^\ast\rangle
  &=&a \left(  \begin{array}{c} -{\rm sin}({{m\pi}\over M}) \cr
                                            {\rm cos}({{m\pi}\over M})
                 \end{array}  \right), \quad \\
\vert\omega_2^\ast\rangle
  &=&b \left(  \begin{array}{c} {\rm sin}({{n\pi}\over M}) \cr
                                            {\rm cos}({{n\pi}\over M})
                 \end{array}  \right),
\end{eqnarray}
\label{Opt_POVM2}
\end{mathletters}
and
\begin{mathletters}
\begin{eqnarray}
a^2&=&{{{\rm cos}({{n\pi}\over M})}\over
            {{\rm sin}({{m\pi}\over M}){\rm sin}({{(m+n)\pi}\over
M})}}\ge0, \\
b^2&=&{{{\rm cos}({{m\pi}\over M})}\over
            {{\rm sin}({{n\pi}\over M}){\rm sin}({{(m+n)\pi}\over
M})}}\ge0.
\end{eqnarray}
Here $m$ and $n$ are any positive integers satisfying
\begin{equation}
0\le a^2+b^2\le2.
\end{equation}
\label{a_b}
\end{mathletters}

\noindent
In some cases one of $a$, $b$ and $\sqrt{2-a^2 -b^2 }$ is zero and
the POVM has only two elements.

{\bf Proof}: For the three element POVM ${\cal W}(\theta , \varphi_a ,
\varphi_b ) =\{\hat\omega_0(\theta), \hat\omega_1(\theta),
\hat\omega_2(\theta)\}$ with rank 1 elements, lemma 6 immediately gives
\begin{eqnarray}
I({\cal E}_M :{\cal W}) &=&  (1-{a^2\over2}-{b^2\over2})I(\theta)
+{a^2\over2}I(\theta+\varphi_a)+{b^2\over2}I(\theta+\varphi_b)
\nonumber
\end{eqnarray}
Hence $ I({\cal E}_M :{\cal W})\leq \max_{\theta} I(\theta )$. By
lemma 3 this maximum is $I({\pi \over 2})$, the accessible
information of ${\cal E}_M$. Furthermore $I(\theta )$ is periodic
in $\theta$ with period ${\pi \over M }$. Hence we can achieve
$I({\cal E}_M : {\cal W}) = I({\pi \over 2})$ by setting
$\theta={\pi \over 2}$ and choosing $\phi_a$ and $\phi_b$ to be any
integer multiples of ${\pi \over M}$. This gives Eqs.
(\ref{Opt_POVM2}). Eqs.  (\ref{a_b}) are just the condition for
$\{\hat\omega_j^\ast\}$ to be a POVM. $\bsquare$

From this theorem we can develop various kinds of optimal
strategies. We noted previously in corollary 1 that if $M$ is even,
then there exists an optimal strategy based on a pair of orthogonal
directions. This also follows from theorem 2: if $M= 4L-2$ with
$L=1,2,\ldots$ then we may take $n=2L-1$ giving $a=0$ and a
2-element POVM based on the directions $\left( \begin{array}{c}
0\\1 \end{array}\right)$ and $\left( \begin{array}{c} 1\\0
\end{array}\right)$. If $M=4L$ with $L=1,2,\ldots$, we may take $m=n=L$
giving $\sqrt{ 2-a^2 -b^2 }=0$ and an optimal POVM based on the
directions $\left( \begin{array}{c} -1\\1 \end{array}\right)$ and
$\left( \begin{array}{c} 1\\1
\end{array}\right)$. In both cases the pair of directions coincides
with an orthogonal pair of states of ${\cal E}_M$.

\begin{figure}
\centerline{\psfig{file=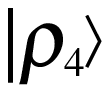,width=5cm}}
\caption{The optimal POVM directions (thick solid lines) given by
theorem 2 in the case of $M=5$. The input states are represented as
$(-1)^k\ket{\psi_k}$ by the dashed lines whose lengths correspond to
a unit state vector.
The lengths of the thick solid lines are scaled according
to the normalization factors of the corresponding POVM elements.}
\label{POVM_5}
\end{figure}

If $M$ is odd, at least 3 outputs are required. In the case of
$M=3$ we get an optimum strategy with three elements of equal norm.
This coincides with our previous result ${\cal A}_3$ of theorem 1
and lemma 3. The cases of $M=5$ and $M=7$ are more interesting. In
both cases, the optimum strategies consist of the three elements
with the two different norms (in contrast to the
$\nums{M}$-covariant strategies of theorem 1). A solution for $M=5$
is shown in Fig. \ref{POVM_5}. The POVM elements are represented by
the thick solid lines and the dashed lines represent the input
states. (Note that, for ease of presentation these dashed lines
representing the states of ${\cal E}_M$
-- symmetrically distributed around a whole circle -- correspond to
the vectors $(-1)^k \ket{\psi_k}$ rather than the original vectors
in Eq. (\ref{psik})). According to choices of parameters $(m,n)$ in
theorem 2, there can be several configurations of the POVM
directions. But by the symmetry of ${\cal E}_5$ they all lie in the
same position relative to the ensemble as a whole, characterized by
$a^2=b^2=1/(2{\rm sin}^2{{2\pi}\over5})$ as shown in Fig.
\ref{POVM_5}. 

\begin{figure}
\centerline{\psfig{file=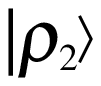,width=10.5cm}}
\caption{The two inequivalent optimal POVMs
in the case of $M=7$. The POVM directions and input states are
represented by thick and dashed lines respectively according to the
conventions of Fig. 2.}
\label{POVM_7}
\end{figure}

Fig. \ref{POVM_7} shows the case of $M=7$. There are
now two inequivalent classes of POVM element directions. One
corresponds to $a^2=b^2=1/(2{\rm sin}^2{{2\pi}\over7})$ where the
angle between the two measurement vectors directed downward is
${2\pi}\over7$ (the left figure), and the other corresponds to
$a^2=b^2=1/(2{\rm sin}^2{{3\pi}\over7})$ where the angle between
the two measurement vectors directed downward is ${3\pi}\over7$
(the right figure).

Lemma 6 and theorem 2 may be used to provide a further variety of
optimal $K$-element POVMs for ${\cal E}_M$ where $K$ is between 3
and $M$:

{\bf Lemma 7:} Let ${\cal A}$ be any POVM as described in lemma 6
for which all angles $\theta_a$ have the form
\begin{equation} \label{tha}
\theta_a = {\pi \over 2} + k_a {\pi \over M}
\hspace{8mm} \mbox{where $k_a$ is an integer}
\end{equation}
Then ${\cal A}$ is an optimal strategy for ${\cal E}_M$.

{\bf Proof:} Since $I(\theta )$ is periodic with period ${\pi \over
M}$ we have $I(\theta_a )=I({ \pi \over 2} )$ for all $a$. Also
$\sum \lambda_a^2 = 2$ so that Eq. (\ref{suma}) immediately gives
$I({\cal E}_M : {\cal A})= I({\pi \over 2 })$ i.e. ${\cal A}$ is
optimal.$\bsquare$

Now note the following facts:\\(a) All POVMs in theorem 2 satisfy
Eq. (\ref{tha}).\\(b) If ${\cal A} = \{ \hat A_i \}$ is any POVM
satisfying Eq. (\ref{tha}) then any $\nums{M}-$shifted version
${\cal A}_l$ of $\cal A$, defined for each $l\in \nums{M}$ by
\[ {\cal A}_l = \{ \hat V^l \hat A_i \hat V^{\dagger l} \} \]
is a POVM also satisfying Eq. (\ref{tha}). (The angles $\theta_a$
are just shifted by ${l\pi \over M}$).\\(c) If ${\cal A}_1, \ldots
,{\cal A}_N$ is any list of POVMs satisfying Eq. (\ref{tha}) then
any convex combination of the ${\cal A}_i$'s will satisfy Eq.
(\ref{tha}). (In forming convex combinations we naturally
amalgamate POVM elements from different ${\cal A}_i$'s that lie in
the same direction.)

Hence any convex combination of any $\nums{M}-$shifted versions of
the POVMs in theorem 2 will be an optimal strategy. For example,
let us consider a convex combination between two POVM's in the case
of $M$=5. The following $\{\hat\omega_j\}$ is one of the optimum
detection strategies from theorem 2:
\begin{mathletters}
\begin{eqnarray}
\hat\omega_0&=&(1-a^2)(\hat I - \hat\sigma_z), \\
\hat\omega_1&=&{a^2\over2}(\hat I
                             - {\rm sin}({{4\pi}\over5})\hat\sigma_x
                             - {\rm cos}({{4\pi}\over5})\hat\sigma_z), \\
\hat\omega_2&=&{a^2\over2}(\hat I
                             - {\rm sin}({{6\pi}\over5})\hat\sigma_x
                             - {\rm cos}({{6\pi}\over5})\hat\sigma_z),
\end{eqnarray}
\end{mathletters}

\noindent
where $a^2=1/(2{\rm sin}^2{{2\pi}\over5})$.
The convex combination between $\{\hat\omega_j\}$ and
$\{\hat V^2 \hat\omega_j\hat V^{\dagger2} \}$ forms the resolution
of the identity
\begin{equation}
(1-\lambda)\sum_{j=0}^{2} \hat\omega_j+ \lambda \sum_{k=0}^{2} \hat
V^2
\hat\omega_k\hat V^{\dagger2}
= \hat I \quad (\lambda\ge0), 
\end{equation}
and we define
\begin{mathletters}
\begin{eqnarray}
\hat\mu_0&=&(1-\lambda)\hat\omega_0
      +\lambda \hat V^2 \hat\omega_2\hat V^{\dagger2}, \\
\hat\mu_1&=&(1-\lambda)\hat\omega_1
      +\lambda \hat V^2 \hat\omega_0\hat V^{\dagger2}, \\
\hat\mu_2&=&(1-\lambda)\hat\omega_2, \\
\hat\mu_3&=&\lambda \hat V^2 \hat\omega_1\hat V^{\dagger2}.
\end{eqnarray}
\label{povm_mu4}
\end{mathletters}

\noindent
(Note that $\hat\omega_0\propto\hat V^2 \hat\omega_2\hat
V^{\dagger2}$ and $\hat\omega_1\propto\hat V^2 \hat\omega_0\hat
V^{\dagger2}$.) This gives a 4-element POVM $\{\hat\mu_j\}$ which
maximizes the Shannon mutual information for ${\cal E}_5$.

The strategies in theorem 2 are not generally $\nums{M}-$covariant
but they correspond to extreme points of $\cal P$. On the other
hand the $\nums{M}-$covariant strategy of theorem 1 is generally
not an extreme point of $\cal P$. The $\nums{M}$-covariant POVM of
theorem 1 can be related to the asymmetrical 3-element POVM of
theorem 2 as follows. Note first that if ${\cal W} =
\{\hat\omega_j\}$ is any optimal POVM then so is $m{\cal W} = \{\hat
V^m\hat\omega_j\hat V^{\dagger m}\} $ for any $m\in
\nums{M}$. Indeed \begin{equation} \label{equal}
I({\cal E}_M : {\cal W})=I({\cal E}_M : m{\cal W})\end{equation}
since the set of states of ${\cal E}_M$ is invariant under the
action of $\nums{M}$. Given any one of the  $N$ (=2, 3)-element POVMs
$\{\hat\omega_j^\ast\}$ defined in theorem 2, one can consider the
resolution of the identity
\begin{equation}
{1\over M}\sum_{m=0}^{M-1}\sum_{j=0}^{N-1}
\hat V^m\hat\omega_j^\ast\hat V^{\dagger m} = \hat I.
\end{equation}
But the $MN$ elements $\{\hat V^m\hat\omega_j^\ast\hat V^{\dagger
m}\}$ are  proportional to each other in groups of $N$ and these
groups may each naturally be summed and assigned a single element.
This leads to the covariant $M$-element POVM which is just ${\cal
A}_M$ of theorem 1 and lemma 3. In this sense ${\cal A}_M$ may be
thought of as a convex combination \[ {\cal A}_M = \sum_{k\in
\nums{M}} {1\over M} k{\cal W} \]
where $\cal W$ is any one of the POVMs in theorem 2. If we know
that ${\cal A}_M$ is optimal then Eqs. (\ref{equal}) and
(\ref{conmax}) will imply that $\cal W$ is optimal too. This
provides an alternative proof of theorem 2 if we already know
theorem 1 and lemma 3. On the other hand, if conversely we are
given the result of theorem 2 (which uses lemma 3) then the
accessible information of ${\cal E}_M$ must be $I({\pi
\over 2})$ so ${\cal A}_M$ must be optimal (since $I({\cal E}_M:
{\cal A}_M )= I({\pi\over 2})$ by definition of $I(\theta )$ and
${\cal A}_M$).

\section{Implementation}

The optimal POVMs ${\cal A}_M$ and ${\cal W}$ given in
theorems 1 and 2, may be of interest from the
viewpoint of putting quantum detection theory to the test.
None of the POVMs for attaining maximum mutual information have
been demonstrated by experiment yet.
So far,
only two kinds of optimal quantum detection scenarios have been
confirmed experimentally. One is the Helstrom bound as the minimum
{\it average} error probability \cite{Helstrom_QDET}, and the other
is the Ivanovic-Dieks-Peres bound which gives the maximum
probability for error-free detection, sometimes referred as the
unambiguous measurement
\cite{Ivanovic87,Dieks88,Peres88,CheflesBarnett98}. (A concise
review of both criteria can be found in ref. \cite{Barnett97RSL}.)
The former scenario was first demonstrated experimentally by
Barnett and Riis \cite{Barnett97_exp}. The latter has been
demonstrated in the laboratory by Huttner et al. \cite{Huttner96}.
Both of these are concerned with discrimination between binary
nonorthogonal states.
In our case of ${\cal A}_M$ and ${\cal W}$ for ${\cal E}_M$ with
$M$ odd, we are dealing with essentially {\it nonorthogonal}
measurement vectors in ${\cal H}_2$, which is called a {\it
generalized} measurement. No von Neumann measurement can be an
optimal strategy for ${\cal E}_M$ with $M$ odd. This case is of
particular interest here. It is already well known that this kind
of generalized measurement can be converted into a standard von
Neumann measurement in a larger Hilbert space by introducing an
ancillary system. This so-called Naimark extension ensures that any
POVM can be physically implemented in principle
\cite{Helstrom_QDET,PeresBook}.

In this section we propose an optical scheme to demonstrate the
optimal POVMs specified by ${\cal W}$ for ${\cal E}_M$ made of
single mode photon polarization states. As seen in the previous
section, ${\cal W}$ has three outcomes at most and suffices to
provide an optimal strategy for all ${\cal E}_M$'s. For $M$ odd, it
is always possible to find the optimal strategy with $m=n$, that
is, $a^2=b^2=1/(2{\rm sin}^2{{m\pi}\over M})$ in theorem 2 if $m$
is taken as ${M\over4}<m<{M\over2}$. We consider the implementation
of this particular detection strategy. The measurement vectors can
be represented by
\begin{mathletters}
\begin{eqnarray}
\vert\omega_0^\ast\rangle&=&
    -{\rm sin}{\gamma\over2}\vert\downarrow\rangle, \\
\vert\omega_1^\ast\rangle&=&
    -{1\over{\sqrt2}}\vert\uparrow\rangle
    +{1\over{\sqrt2}}{\rm cos}{\gamma\over2}
                     \vert\downarrow\rangle, \\
\vert\omega_2^\ast\rangle&=&
     {1\over{\sqrt2}}\vert\uparrow\rangle
    +{1\over{\sqrt2}}{\rm cos}{\gamma\over2}
                     \vert\downarrow\rangle,
\end{eqnarray}
\label{vec_omega_m=n}
\end{mathletters}

\noindent
where
\begin{equation}
{\rm cos}{\gamma\over2}\equiv{\rm cot}{{m\pi}\over M},
\quad
{\rm sin}{\gamma\over2}\equiv
-\sqrt{1-{\rm cot}^2{{m\pi}\over M}},
\label{angle_gamma}
\end{equation}
and $\vert\uparrow\rangle$ and $\vert\downarrow\rangle$ are
orthonormal bases of polarization.
The first step is to make orthogonal measurement vectors by
embedding
$\{\vert\omega_0^\ast\rangle,
   \vert\omega_1^\ast\rangle,
   \vert\omega_2^\ast\rangle\}$
into a three or higher dimensional Hilbert space. One possible
physical prescription is to make an optical circuit with two input
ports, say, ``a" and ``b". The signal state is guided into the port
``a", while the port ``b" is initialized as the vacuum state. We
can then consider the four dimensional Hilbert space spanned by the
orthonormal basis $\{\vert E_j\rangle\}$,
\begin{mathletters}
\begin{eqnarray}
\vert E_0\rangle&\equiv&\vert\uparrow\rangle_a \vert0\rangle_b, \\
\vert E_1\rangle&\equiv&\vert\downarrow\rangle_a \vert0\rangle_b,\\
\vert E_2\rangle&\equiv&\vert0\rangle_a \vert\uparrow\rangle_b, \\
\vert E_3\rangle&\equiv&\vert0\rangle_a \vert\downarrow\rangle_b,
\end{eqnarray}
\label{basis_E}
\end{mathletters}
where $\vert0\rangle$ is the vacuum state and the subscripts $a$ and $b$
indicate the port $``a"$ and $``b"$, respectively.
A natural orthogonalization is
\begin{mathletters}
\begin{eqnarray}
\vert\Omega_0\rangle  &\equiv&
       \vert\omega_0^\ast\rangle_a \vert0\rangle_b
    + {\rm cos}{\gamma\over2}\vert0\rangle_a \vert\uparrow\rangle_b, \\
\vert\Omega_1\rangle  &\equiv&
       \vert\omega_1^\ast\rangle_a \vert0\rangle_b
    + {1\over{\sqrt2}}{\rm sin}{\gamma\over2}
       \vert0\rangle_a \vert\uparrow\rangle_b, \\
\vert\Omega_2\rangle  &\equiv&
       \vert\omega_2^\ast\rangle_a \vert0\rangle_b
    + {1\over{\sqrt2}}{\rm sin}{\gamma\over2}
      \vert0\rangle_a \vert\uparrow\rangle_b, \\
\vert\Omega_3\rangle  &\equiv&
       \vert0\rangle_a \vert\downarrow\rangle_b,
\end{eqnarray}
\label{vec_Omega_1}
\end{mathletters}
or equivalently,
\begin{mathletters}
\begin{eqnarray}
\vert\Omega_0\rangle  &\equiv&
    - {\rm sin}{\gamma\over2}\vert\downarrow\rangle_a \vert0\rangle_b
    + {\rm cos}{\gamma\over2}\vert0\rangle_a \vert\uparrow\rangle_b, \\
\vert\Omega_1\rangle  &\equiv&
      {1\over{\sqrt2}}\left(
    - \vert\uparrow\rangle_a \vert0\rangle_b
    + {\rm cos}{\gamma\over2}\vert\downarrow\rangle_a \vert0\rangle_b
    + {\rm sin}{\gamma\over2}
       \vert0\rangle_a \vert\uparrow\rangle_b\right), \\
\vert\Omega_2\rangle  &\equiv&
      {1\over{\sqrt2}}\left(
      \vert\uparrow\rangle_a \vert0\rangle_b
    + {\rm cos}{\gamma\over2}\vert\downarrow\rangle_a \vert0\rangle_b
    + {\rm sin}{\gamma\over2}
       \vert0\rangle_a \vert\uparrow\rangle_b\right), \\
\vert\Omega_3\rangle  &\equiv&
       \vert0\rangle_a \vert\downarrow\rangle_b.
\end{eqnarray}
\label{vec_Omega_2}
\end{mathletters}

\begin{figure}
\centerline{\psfig{file=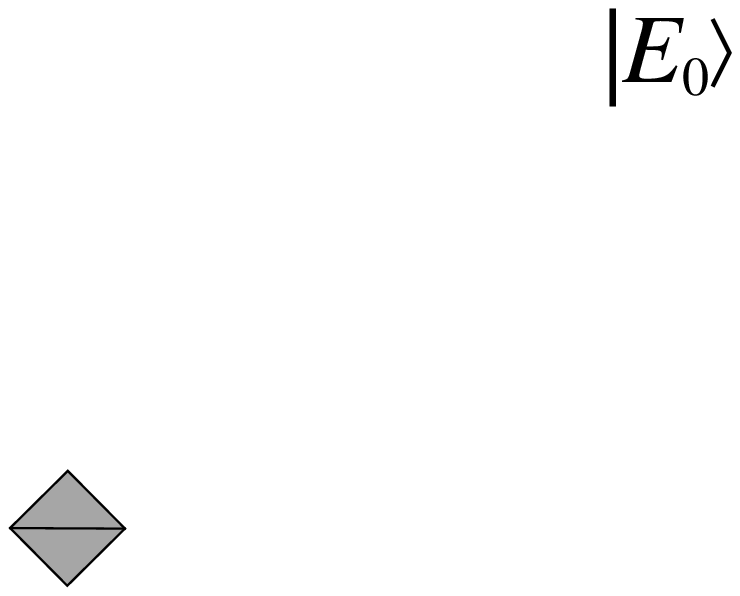,width=10cm}}
\caption{The optical circuit implementing
${\cal W}=\{\hat\omega_0^\ast,\hat\omega_1^\ast,\hat\omega_2^\ast\}$.
It consists of the unitary transformation $\hat U_2\hat U_1$
followed by the measurement $\{\vert E_j\rangle\}$.
$\hat U_2\hat U_1$ is effected by four halfwave plates, two
polarizing beam splitters, and two polarization rotators.
The measurement $\{\vert E_j\rangle\}$ is made by photon counting
at the four output ports.}
\label{Decoding_circuit}
\end{figure}

\noindent
It is easy to check that
$\{\vert\Omega_0\rangle,\vert\Omega_1\rangle,\vert\Omega_2\rangle\}$
give the same channel matrix as
$\{\vert\omega_0^\ast\rangle,\vert\omega_1^\ast\rangle,
\vert\omega_2^\ast\rangle\}$, that is,
$\langle\omega_j\vert\psi_i\rangle
=\langle\Omega_j\vert
(\vert\psi_i\rangle_a\vert0\rangle_b)$ $(j=0,1,2)$. The second step
is to decompose the von Neumann measurement
$\{\vert\Omega_j\rangle\}$ into a unitary transformation followed
by a measurement in the basis $\{\vert E_j\rangle\}$ in order to
find a practical detector structure. We may write
\begin{mathletters}
\begin{eqnarray}
\langle\Omega_0\vert&\equiv&\langle E_2\vert\hat U_2\hat U_1, \\
\langle\Omega_1\vert&\equiv&\langle E_1\vert\hat U_2\hat U_1, \\
\langle\Omega_2\vert&\equiv&\langle E_0\vert\hat U_2\hat U_1, \\
\langle\Omega_3\vert&\equiv&\langle E_3\vert\hat U_2\hat U_1,
\end{eqnarray}
\label{U+E}
\end{mathletters}
where $\hat U_1$ and $\hat U_2$ are given by the matrices
\begin{equation}
\hat U_1\equiv
\left(  \begin{array}{cccc}
1&                       0&                      0&0 \cr
0& {\rm cos}{\gamma\over2}&{\rm sin}{\gamma\over2}&0 \cr
0&-{\rm sin}{\gamma\over2}&{\rm cos}{\gamma\over2}&0 \cr
0&                       0&                      0&1
        \end{array}  \right),
\label{U_1}
\end{equation}
\begin{equation}
\hat U_2 \equiv
\left(  \begin{array}{cccc}
 {1\over\sqrt{2}}&{1\over\sqrt{2}}&0&0 \cr
-{1\over\sqrt{2}}&{1\over\sqrt{2}}&0&0 \cr
                0&               0&1&0 \cr
                0&               0&0&1
           \end{array}  \right),
\label{U_2}
\end{equation}
in the
$\{\vert E_0\rangle,\vert E_1\rangle,\vert E_2\rangle,\vert E_3\rangle\}$
-basis representation. Eqs. (\ref{U+E}) mean that in the detector,
the signal state $\vert\psi_i\rangle_a\vert0\rangle_b$ is first
transformed by $\hat U_2\hat U_1$, and is then measured in the
basis $\{\vert E_j\rangle\}$ which corresponds to the simultaneous
measurement with respect to {\it which-path} and {\it
which-polarization}. The final step is to translate $\hat U_2\hat
U_1$ into a practical circuit. In fact, this unitary transformation
can be effected by the simple circuit consisting of passive linear
optical devices such as polarizing beam splitters, polarization
rotators, and halfwave plates \cite{Cerf98}. The circuit is shown
in Fig. \ref{Decoding_circuit}. 
The $\hat U_2\hat U_1$ part
consists of four halfwave plates, two polarizing beam splitters,
and two polarization rotators. The polarization rotator represented
by the circle with the rotation angle $\gamma$ performs
\begin{equation}
\hat R_y(\gamma)=
\left(  \begin{array}{cc}
{\rm cos}{\gamma\over2}&{\rm sin}{\gamma\over2} \cr
-{\rm sin}{\gamma\over2}&{\rm cos}{\gamma\over2}
           \end{array}    \right).
\label{R_y}
\end{equation}
The polarizing beam splitter represented by the square functions 
as a perfect mirror only for $\downarrow$-polarization 
(fast axis polarization). Light polarized along 
$\uparrow$-polarization (slow axis polarization) passes straight 
through it perfectly. 
The measurement $\{\vert E_j\rangle\}$ is made by photon counting
at the four output ports. Note that only a single photon count at
one of the three ports is expected and the outcome $\vert
E_3\rangle$ is never expected. This structure is valid for any $M$
(the number of the signals) if one tunes the rotation angle
$\gamma$ in $\hat R_y(\gamma)$ according to the value of $M$ (see
Eq. (\ref{angle_gamma})). The circuit is simple enough to be
implemented with present technology.

\section{Concluding remarks}\label{sec5}

We have considered optimal strategies for symmetrical sources of
real quantum states, treating in detail the sources ${\cal E}_M$ of
$M$ real qubit states placed symmetrically in the $x-z$ plane
around the Bloch sphere. Davies \cite{Davies78} has provided a
general theorem characterising an optimal strategy for any
$G$-covariant source whose group acts irreducibly on the whole
state space. The symmetry group $\nums{M}$ of ${\cal E}_M$ does not
act irreducibly on that state space so Davies' theorem cannot be
directly applied. However we proved an extension of this theorem
which applies to $G$-covariant sources of real states for which the
group acts irreducibly on the subset of real states (as is the case
for ${\cal E}_M$). This led to a $\nums{M}$-covariant optimal
strategy ${\cal A}_M$ for ${\cal E}_M$.

We also derived alternative optimal strategies ${\cal W}$ which
contain at most three real POVM elements. In deriving this strategy
${\cal W}$ we exploited the convexity of $I(X:Y)$ on the convex set
${\cal P}$ of all POVMs. These strategies are not $G$-covariant in
general but correspond to extreme points of ${\cal P}$. The small
number of elements can be advantageous for practical implementation
of the detection strategies as seen in the preceding section. The
$G$-covariant strategy is not generally an extreme point of ${\cal
P}$ but for higher dimensions it would seem easier to derive
explicit $G$-covariant solutions rather than extreme point
solutions.

Our results have added to the relatively small number of quantum
sources for which optimal strategies are explicitly known. They may
be extended in various straightforward ways (which we have omitted
for clarity of presentation). For example the optimal strategies
${\cal A}_M$ and ${\cal W}$ for ${\cal E}_M$ remains optimal
for the $M$-state source
\[
\{ (1-\epsilon ) \proj{\psi_k} + \epsilon {1\over 2}\hat I_2 
 : k\in \nums{M}; {1\over M} \} \]
where each pure signal has been corrupted by noise given by the
maximally mixed state ${1\over 2}\hat I_2$. 
This mixed state ensemble
is clearly also $G$-covariant and the process of deriving the
optimal strategy for this ensemble is quite the same as in the
pure state case ($\epsilon=0$) but just multiplying the cosine
terms in Eq. (\ref{mutinf}) by $(1-\epsilon )$. Then the same
strategy remains optimal for the $G$-covariant mixed state ensemble
although the accessible information decreases with $\epsilon$ as
expected.

It is perhaps worth briefly contrasting our results of maximizing
the mutual information with the problem of minimizing the average
error probability. The latter is defined for ${\cal E}_M$ and any
$M$-element POVM by
\begin{equation}
P_{\rm e}= 1- { 1\over M}\sum_{k=0}^{M-1}P(k\vert k).
\label{Pe}
\end{equation}
The $P_{\rm e}$-optimal strategy is $\{ \hat\pi_k \} = \{ {2\over
M}
\proj{\psi_k}:k\in\nums{M} \}$, that is, the POVM based on the
state directions themselves. This is true also for the above mixed
state ensemble. (The necessary and sufficient conditions for
$P_e$-optimality, as given in
\cite{Holevo73_condition,Helstrom_QDET}, are easily verified for
$\{\hat\pi_k\}$.) Generally $P_e$-minimization is an essentially
different type of optimization problem from $I(X:Y)$-maximization.

Within the confines of our formalism, various interesting issues
remain unresolved. For example we would like to know an optimal
strategy for the real $\nums{M}$-covariant source ``double-${\cal
E}_M$'' in 4 dimensions comprising the 2-qubit signal states $\{
\ket{\psi_k}\ket{\psi_k} : k\in \nums{M} ; {1\over M} \} $. In this
case the symmetry group $\nums{M}$ does not act irreducibly even on
the subset of all real 2-qubit states. Interesting properties of
double-${\cal E}_3$ have been considered in \cite{pw92} from the
viewpoint of coding gain of transmittable information.

It is also a remaining difficult problem to optimize a quantum
channel over both the {\em a priori} probability distribution of
signals {\it and} the detection strategy for a fixed set of quantum
states. The solution is known only for the binary pure state
channel.

\acknowledgements

The authors would like to thank A. S. Holevo and T. S. Usuda
for giving crucial comments on this work.
They would also like to thank C. A. Fuchs, C. H. Bennett, and
A. Chefles for helpful discussions.
RJ is grateful to W. K. Wootters who in 1993 drew his attention
to the reality of the source ${\cal E}_3$ as an important property,
which ultimately led to theorem 1.
MS and SMB thank the Great Britain Sasakawa Foundation and
the British Council for financial support.
SMB and RJ thank the UK Engineering and Physical
Science Research Council for financial support.

\end{document}